\documentclass[a4paper,fleqn,usenatbib,11pt]{mnras2}

\usepackage[T1]{fontenc}
\usepackage{ae,aecompl}

\usepackage{array,url}

\usepackage{graphicx}	% Including figure files
\usepackage{amsmath}	% Advanced maths commands
\usepackage{amssymb}	% Extra maths symbols

\usepackage{url}
\usepackage{graphicx}
\newcommand{\be}{\begin{equation}}
\newcommand{\ee}{\end{equation}}
\newcommand{\bea}{\begin{eqnarray}}
\newcommand{\eea}{\end{eqnarray}}

\def\arcsec{^{\prime\prime}}
\def\arcmin{^\prime}

\usepackage{color}              % Need the color package
\usepackage{epsfig}
\usepackage{amstext}

\vspace{-5em}
\title[Non--thermal SZ effect from galaxy jey]{Relativistic inverse Compton scattering of photons from the early universe}
\author[Malu, Datta, Colafrancesco, Marchegiani et al.]{
S. Malu$^{1}$\thanks{E-mail: siddharth@iiti.ac.in},
A. Datta$^{1,3}$,
S. Colafrancesco$^{2}$,
P. Marchegiani$^{2}$,
R. Subrahmanyan$^{4}$, \\ \Large \normalfont
D. Narasimha$^{5}$,
M. H. Wieringa$^{6}$
\\
$^{1}$Centre of Astronomy, Indian Institute of Technology Indore, Simrol, Khandwa Road, Indore 453552, India \\
$^{2}$School of Physics, University of the Witwatersrand, Private Bag 3, WITS-2050, Johannesburg, South Africa\\
$^{3}$Center for Astrophysics and Space Astronomy, Department of Astrophysical and Planetary Science, University of Colorado, \\ 
Boulder, CO 80309, USA\\
$^{4}$Raman Research Institute, CV Raman Avenue, Bangalore 560080, India\\ 
$^{5}$Tata Institute of Fundamental Research, Homi Bhabha Road,
Mumbai 400005, India \\ 
$^{6}$Australia Telescope National Facility, CSIRO Astronomy and Space Science, P.O. Box 76, Epping NSW 1710
}

\begin{document}
\label{firstpage}
\pagerange{\pageref{firstpage}--\pageref{lastpage}}
\maketitle
%\twocolumn[{%
% \centering
% \Large {\bf Relativistic inverse Compton scattering of photons from the early universe} \\
% \large Siddharth Malu\Mark{$\ast$1},
 %       Abhirup Datta\Mark{1,3},
  %      Sergio Colafrancesco\Mark{2},
   %     Paolo Marchegiani\Mark{2},
   %     Ravi Subrahmanyan\Mark{4},
     %   D Narasimha\Mark{5},
       % Mark H. Wieringa\Mark{6}\\
 %\normalsize
%% \begin{tabular}{*{2}{>{\centering}p{.25\textwidth}}}
 % \Mark{1}Centre of Astronomy, Indian Institute of Technology Indore,
%Simrol, Khandwa Road, Indore 453552, India \\ \Mark{2}School of
%Physics, University of the Witwatersrand, Private Bag 3, WITS-2050,
%Johannesburg, South Africa \\ \Mark{3}Center for Astrophysics and
%Space Astronomy, Department of Astrophysical and Planetary Science,
%University of Colorado, Boulder, CO 80309, USA \\ \Mark{4}Raman
%Research Institute, CV Raman Avenue, Bangalore 560080, India
%\\ \Mark{5}Tata Institute of Fundamental Research, Homi Bhabha Road,
%Mumbai 400005, India \\ \Mark{4}Australia Telescope National Facility,
%CSIRO Astronomy and Space Science, P.O. Box 76, Epping NSW 1710 \\ 
%  $^{\ast}$\url{siddharth@iiti.ac.in}
% \end{tabular}\\ % some more space after the title part
%}]
%\normalsize
%{\huge \bf Probing non--equilibrium particle distribution through CMB Spectral Distortions}

%% Abstract of the paper
\begin{abstract}
%{\bf 
Electrons at relativistic speeds, diffusing in magnetic fields, cause
  copious emission at radio frequencies in both 
  clusters of galaxies and radio galaxies, through the non-thermal radiation emission called synchrotron. 
  However, the total power radiated through this mechanism is ill constrained,
  as the lower limit of the electron energy distribution, or low--energy cutoffs, for radio emission in galaxy clusters and radio   
  galaxies have not yet been determined. This lower limit, parametrized by the
  lower limit of the electron momentum -- p$_{\rm min}$ -- is critical for
  estimating the energetics of non-thermal electrons produced by cluster mergers or injected by radio galaxy jets, which  
  impacts the formation of large--scale structure in the universe, as well as the
  evolution of local structures inside galaxy clusters.\\
  The total pressure due to the relativistic, non--thermal population of electrons is critically
  dependent on p$_{\rm min}$, making the measurement of this non--thermal
  pressure a promising technique to estimate the electron low--energy
  cutoff. We present here the first unambiguous detection of this
  pressure for a non--thermal population of electrons in a radio galaxy
  jet/lobe, located at a significant distance away from the center of the
  Bullet cluster of galaxies.
%}\\
\end{abstract}
\section{Introduction}
\label{intro}
Galaxy clusters, the largest gravitationally bound structures in
  the universe, host large populations of hot plasma -- electrons -- at temperatures of
  up to 10--100 million Kelvins. 
  A fraction of these galaxy clusters also host energetic population of relativistic electron plasma whose nature is still
   unknown. Possible sources have been proposed in relation with the existence of intra-cluster shock waves, the injection
   processes from radio and active galaxies and dark matter annihilation/decay.
Once these relativistic electrons are produced by one or a combination
of these mechanisms, they diffuse in the magnetized cluster atmosphere
and emit non-thermal synchrotron radio emission observed as both
extended arches and filamentary structures (named radio relics) or in
more homogeneously diffuse halos (name radio halos). 
  
%  Galaxy clusters grow through mergers \textbf{and accretion} with other clusters and galaxy
%  groups, and the mergers create shock waves in the intracluster gas,
%  which accelerates particles to relativistic speeds.\\
%  A significant/critical issue in the study of shocks in galaxy
%  cluster mergers through synchrotron emission is the energy
%  distribution of these accelerated particles. 
A significant/critical issue in the study of non-thermal electrons through synchrotron emission is the energy
  distribution of these accelerated particles.
  Basic parameters that characterize the energy distribution of the
  electrons in the atmospheres of the galaxy cluster are the high- and
  low--energy cutoffs.\\
  Since the energy spectrum of synchrotron emission produced by the accelerated
  electrons in clusters is expected to follow a power law whose intensity decreases with increasing energy/frequency,
  therefore, it is the low--energy cutoff that is crucial for
  determining the total energetics of the cluster--wide radio emission.\\
  Radio emission from jets of radio galaxies is also due to synchrotron emission of relativistic electrons, and
  therefore also has a power law, and the low--energy cutoff is
  similarly a critical quantity to determine the energetics of the radio jets
  associated with these radio galaxies.\\ 
  Both thermal and non--thermal populations of electrons in galaxy
  clusters, as well as the mostly non--thermal populations of
  electrons in the jets of radio galaxies, are energetic enough to
  cause photons, traveling from the early universe (the Cosmic
  Microwave Background or CMB), to shift to higher energies, hence shifting the
  entire CMB spectrum of photons. This is due to the
    fundamental mechanisms of inverse Compton scattering (ICS) and it
    is usually referred to as the Sunyaev--Zel'dovich Effect (or SZ
  Effect) for the up-scattering produced by thermal
  populations of electrons. Due to the universality of the ICS
    mechanisms, also non-thermal and relativistic electrons in
    galaxy clusters can up-scatter the CMB photons by largely increasing
    their final frequency. This leads to a more general form of the SZ
    Effect that we refer to as non-thermal SZE for simplicity.\\  
  The SZ Effect probes the integrated pressure (or energy
    density) of the relative electron population (thermal or
    non-thermal) in galaxy clusters, along the line of sight -- and this property renders it
  a critical probe of the plasma in these cosmic structures, since it yields
  information complementary to radio emission from synchrotron. While
  radio emission from synchrotron provides information about the
  presence of non-thermal electrons embedded in magnetic fields, the SZ Effect
  provides a ``snapshot'' of their pressure profiles; in particular,
  pressure enhancements, along shocks, radio galaxy jets and other regions in
  galaxy clusters; especially mergers or collisions of these galaxy clusters.\\
  
%  However, the most interesting physics, i.e. the information about
%  the low--energy cutoff for the energy distribution of %electrons,
%  lies in non--equilibrium 
%  portions of the cluster and galaxy radio jet plasma. Hence the need
%  to study spectral distortions produced by populations of electrons that are out
%  of equilibrium in galaxy clusters -- this is known as the `non--thermal SZ
%  effect' or ntSZ. \\
%We present here the first unambiguous detection of the non--thermal SZ Effect  in a radio galaxy
%jet/lobe, which is $\sim$ 800 kpc away from the center of the Bullet
%cluster of galaxies.\\ 
%
%\begin{figure}
%%  \begin{center}
%	\includegraphics[width=\columnwidth,angle=0]{fig1try2.eps}
%    \caption{The `Bullet' cluster of galaxies, which is a cluster
%      collision/merger, from \citet{2006ApJ...648L.109C}. This figure
%      is an overlay of X-ray (red) -- which traces gas at equilibrium -- and
%      weak lensing (blue) data (weak lensing being proxy for Dark
%      Matter; see \citet{2006ApJ...648L.109C} for a detailed
%      discussion). The approximate position of the radio galaxy RG01,
%      including its jet, has been marked as a white shape on the
%      figure, and labeled.} 
%    \label{fig1}
%%  \end{center}
%\end{figure}
\section{The `Bullet' cluster of galaxies}
A spectacular example of an extremely energetic merger or collision of
clusters of galaxies is the `Bullet' cluster (1E0657$-$56) , a southern sky object,
named due to the eponymous shape of the smaller cluster. This cluster merger
provided the most direct evidence for the existence of the so--called
Dark Matter \citep{2006ApJ...648L.109C}, which was found to be
significantly displaced with respect to X--ray emission -- this
is also one of the most X-ray luminous clusters observed.
%It is one of the hottest known
%clusters, has been well-studied over the last decade for a variety
%of reasons; namely, %its strong cluster collision/merger event at
%$z\sim$0.296, with the larger and more massive, eastward cluster being
%$\sim$10 times the mass of the smaller `bullet', 
Other reasons that make this cluster rich in non--equilibrium physics,
and one of the most interesting objects to study, are: the existence of a strong radio halo \citep{2000ApJ...544..686L,2014MNRAS.440.2901S}
which can be observed up to cm--wavelengths
\citep{2011JApA...32..541M}, a bright radio relic
\cite{2015MNRAS.449.1486S} which has been observed up to 10 GHz
\citep{2016Ap&SS.361..255M,2015MNRAS.449.1486S}, and the presence of a
thermal SZE (\citet{2009ApJ...701...42H}, \citet{2010ApJ...716.1118P}
and references therein).%, its peculiar X-ray brightness distribution
%\citep{2002ApJ...567L..27M}, %though most notably in providing the
                             %most direct proof of the existence of
                             %dark matter \citep{2006ApJ...648L.109C}
                             %which is offset w.r.t. the
                             %X-rayemission. 
%The Bullet cluster is also one of the most X-ray luminous galaxy
%clusters, with a bow--shock structure, and cluster--wide extended emission from the radio halo detected up to
%18 GHz \citep{2011JApA...32..541M}, and a radio relic emission
%detected up to 10 GHz \citep{2016Ap&SS.361..255M,2015MNRAS.449.1486S}.\\
%
We present here the first unambiguous detection of the non--thermal SZ
Effect in a radio galaxy jet/lobe, which is $\sim$ 800 kpc away from the center of the Bullet
cluster of galaxies. Importantly, we use low-frequency radio data in the range (2.1-9.0) GHz to determine the
spectrum of the radio lobe and then fit the ATCA 18 GHz SZE
observation with a non-thermal SZE model that is computed in a fully
relativistic approach.\\ 
Throughout the paper, we use a flat, vacuum--dominated cosmological model with $\Omega_m = 0.315$, $\Omega_{\Lambda} = 0.685$ and $H_0 =67.3$ km s$^{-1}$ Mpc$^{-1}$.

\section{The radio galaxy RG01}
%{\large \bf The radio galaxy RG01}\\
%\section{The radio galaxy RG01}
The Bullet cluster was observed using the Australia Telescope Compact
Array (ATCA) at 18 GHz center frequency (16--20 GHz range), using the
two most compact arrays, H75 and H168. Details of the observations
that are used to image the SZE are provided in \citet{2011JApA...32..541M}, and in Table
\ref{obs_journal}.\\ 
\begin{center}
\begin{table}
\caption{Summary of the ATCA observations}
\label{obs_journal}
\begin{tabular}{@{}lr}
\hline\hline
Co-ordinates (J2000) RA -- Dec & $06^{\rm h}58^{\rm m}14.3^{\rm s} -55\degr54\arcmin24\arcsec$ \\ 
Primary Beam FWHM & 2.6$\arcmin$ \\
Synthesized Beam FWHM: & \\
22.6$\arcsec\times$15.9$\arcsec$ Natural Weighting & (Without Antenna 6) \\
15.7$\arcsec\times$12.3$\arcsec$ Natural Weighting & (With Antenna 6) \\
7.7$\arcsec\times$ 5.9$\arcsec$ Uniform Weighting & (With Antenna 6) \\
19.5$\arcsec\times$13.8$\arcsec$ Uniform Weighting & (Without Antenna 6) \\
Frequency Range & 16--20 GHz \\
Total observing time & 140 hours \\
Arrays & H168 \& H75 \\
Amplitude Calibrator & PKS B1934$-$638 \\
Phase Calibrator & PKS~B0742$-$56 \\
Frequency Resolution & 1 MHz \\
%Expected Sensitivity & 1.4$\mu$Jy beam$^{-1}$\\
Sensitivity & 4.0 $\mu$Jy beam$^{-1}$\\
Pointings & 2 (2009 observations) \\
          & 4 (2010 observations) \\
\hline\hline
\end{tabular}
\end{table}
\end{center}

%\begin{figure}
%%  \begin{center}
%	\includegraphics[width=\columnwidth,angle=0]{4freqs2.eps}
%    \caption{{\bf (a) -- (c) \& (e)}: The radio galaxy RG01 at 2.1,
%      9.0 and 18 GHz. {\bf (d)}: DSS image of RG01, which is clearly pointed out as the central, faint source.} 
%    \label{4freqs}
%%  \end{center}
%\end{figure}
%\vspace{-4em}
RG01 is one of the radio galaxies detected in the Bullet cluster field
at frequencies 2.1, 5.5, 9.0 and 16--24 GHz in ATCA observations.% --
%its position is reported in Table \ref{obs_journal}. %Images of the
%radio galaxy at the three radio frequencies 2.1, 5.5 \& 9.0 GHz, and a
%\textbf{DSS-2 red optical} image, are presented in Fig. \ref{4freqs}.\\
%
The radio galaxy RG01 -- and therefore its radio jet/lobe -- is
located approximately 180$\arcsec$ or $\approx$ 800 kpc away from the
center of the Bullet cluster. The radio galaxy, and the jet/lobe
region, is also approximately 160$\arcsec$, or $\approx$ 700 kpc away
from the nearest radio halo region with diffuse radio emission, which
is detected up to 10 GHz
\citep{2016Ap&SS.361..255M,2015MNRAS.449.1486S}. In addition, there is
no detectable X-ray emission with Chandra in this region. These facts (i.e., the large distance between RG01 and
the center of the Bullet cluster, the absence of X-ray emission, the
region containing RG01 being far away from the shock front as well as
the cold front, and the SZE being in the radio jet/lobe region where
synchrotron has already been detected at 2.1, 5.5 and 9.0 GHz) imply
that the SZE we detect can only be of non-thermal origin.\\

\subsection{The non-thermal SZE in the RG01 lobe}
\begin{figure*}
%  \begin{center}
	\includegraphics[width=1.01\textwidth,angle=0]{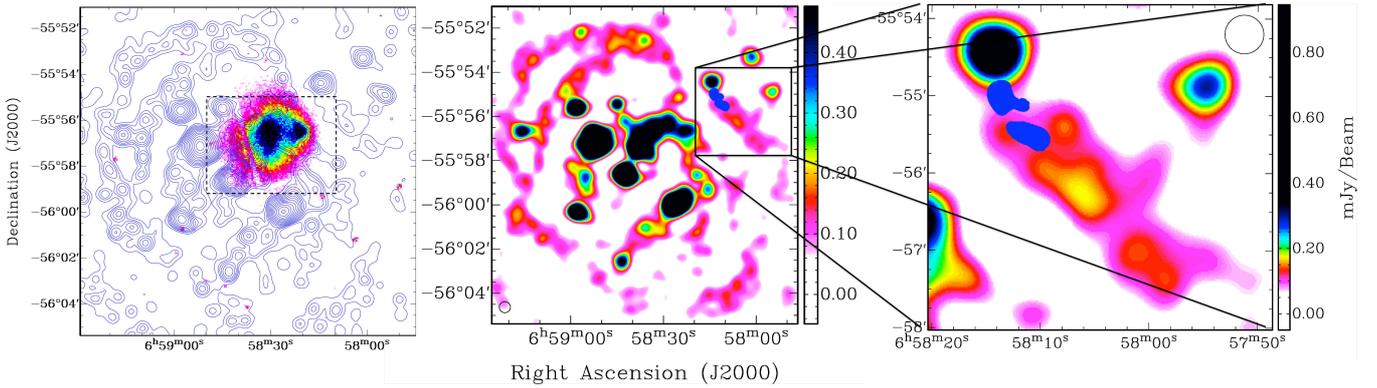}
    \caption{{\bf Left}: 5.5 GHz contours on X--ray colour, showing the relative
    position of RG01 and the X--ray emission in the Bullet
    cluster. X--ray data was obtained from the Chandra Data Archive (500ks observations described in \citet{2006ESASP.604..723M}), and was displayed using the KARMA package \citep{1996ASPC..101...80G}. Radio contour levels start at 5$\sigma$ and increase by a
    factor of $\sqrt{2}$. The position and relative size of the image in Fig. 1 of \citet{2006ApJ...648L.109C}
    have been indicated and marked. This comparison clearly
    indicates the significantly larger spread of non--thermal emission
    in radio as compared to the thermal X--ray emission. {\bf Middle}: the non-thermal SZE detected in the radio galaxy lobe/jet at
      18 GHz, displayed as a blue region in the NW region of the 5.5 GHz
      ATCA image, with 30$\arcsec$ resolution. The  SZE
      region in blue is from an image with the same resolution of
      30$\arcsec$, having been convolved with a 30$\arcsec$ beam. 
      The 5.5 GHz image has a noise rms of 14$\mu$Jy
      beam$^{-1}$, and the 18 GHz image has a noise rms of 3.5$\mu$Jy
      beam$^{-1}$. A 30$\arcsec$ beam size is represented in the bottom
      left corner. {\bf Right}: a zoom--in to the region marked with a rectangle on
      the left image; the non-thermal SZE is marked in blue, and two distinct regions can
      clearly be seen, one closer to the radio galaxy core than the
      other. The 30 $\arcsec$ beam is marked in the top right
      corner. The colour scheme for the 5.5 GHz image is somewhat
      different, to accentuate small-scale features at 5.5 GHz. To the
      NW of the radio galaxy and radio lobe/jet there is another radio galaxy,
      and to the SE is the NW tip of the radio halo at 5.5 GHz, which
      helps mark the location of the non-thermal SZE in relation to the radio halo.}
    \label{composite01}
%  \end{center}
\end{figure*}

We detected a non-thermal SZE signal in the jet/lobe of the radio galaxy RG01 located at coordinates (J2000)RA: $06^{\rm
  h}58^{\rm m}14.2^{\rm s}$ DEC: $-55\degr54\arcmin25\arcsec$ and
shown in the blue-colored region of Fig. \ref{composite01}. Given the
noise rms of 4 $\mu$Jy beam$^{-1}$, this is a 5.5$\sigma$ detection,
with the deepest SZE signal being $-$22$\mu$Jy beam$^{-1}$. We
produced images with the five different values of the FWHM,%(four
%indicated in Table \ref{obs_journal}, and the 30$\arcsec$ being the
%fifth one)
with different weightings, and different amounts of
uv--coverages, as reported in Table \ref{obs_journal}. 
With these different weighting schemes, as well as different FWHMs of synthesized
beams, the size of the two SZE regions turns out to be at most 5\%
different and thus retain the same morphology. Given that the natural
and uniform weighing schemes provide different/independent beams, any
detection above 3--5$\sigma$ that does not depend on the weighing
scheme (and therefore on the deconvolution process), is therefore
considered as a signal. This demonstrates that our detection of the non-thermal
SZE in this region is robust. Additionally, the effect of the
synthesized side-lobes is at maximum of 2\% at these angular distances
away from the brightest sources, assuming that it is 3$\arcmin$
away.
Conversion from brightness to temperature units is given by $T =
\frac{\lambda^{2}}{2k\Omega}S$ and, given that $\lambda=$1.67 cm
(i.e., 18 GHz), we obtain $T = 1.19 \times B$ where $B$ is in units of
$\frac{\rm{Jy}}{\rm{arcmin}^{2}}$. Since the non-thermal SZE we have shown is
taken from an image with a beam of 30$\arcsec\times$30$\arcsec$, our
detection of this SZE at $-$22$\mu$Jy beam$^{-1}$ corresponds,
therefore, to $-$88$\mu$Jy arcmin$^{-2}$; from the formula in the
previous equation this yields $\Delta T_{\rm{SZ}}=-105\mu K$, and a Compton-y parameter of 
%\begin{equation}
$y=-\frac{1}{2}\frac{\Delta T_{\rm{SZ}}}{T_{\rm{CMB}}}=1.9\times10^{-5}$.\\
%  \label{eq04}
%\end{equation}
%
\section{Modeling the non--thermal SZ Effect}
\begin{figure}
%  \begin{center}
	\includegraphics[width=\columnwidth,angle=0]{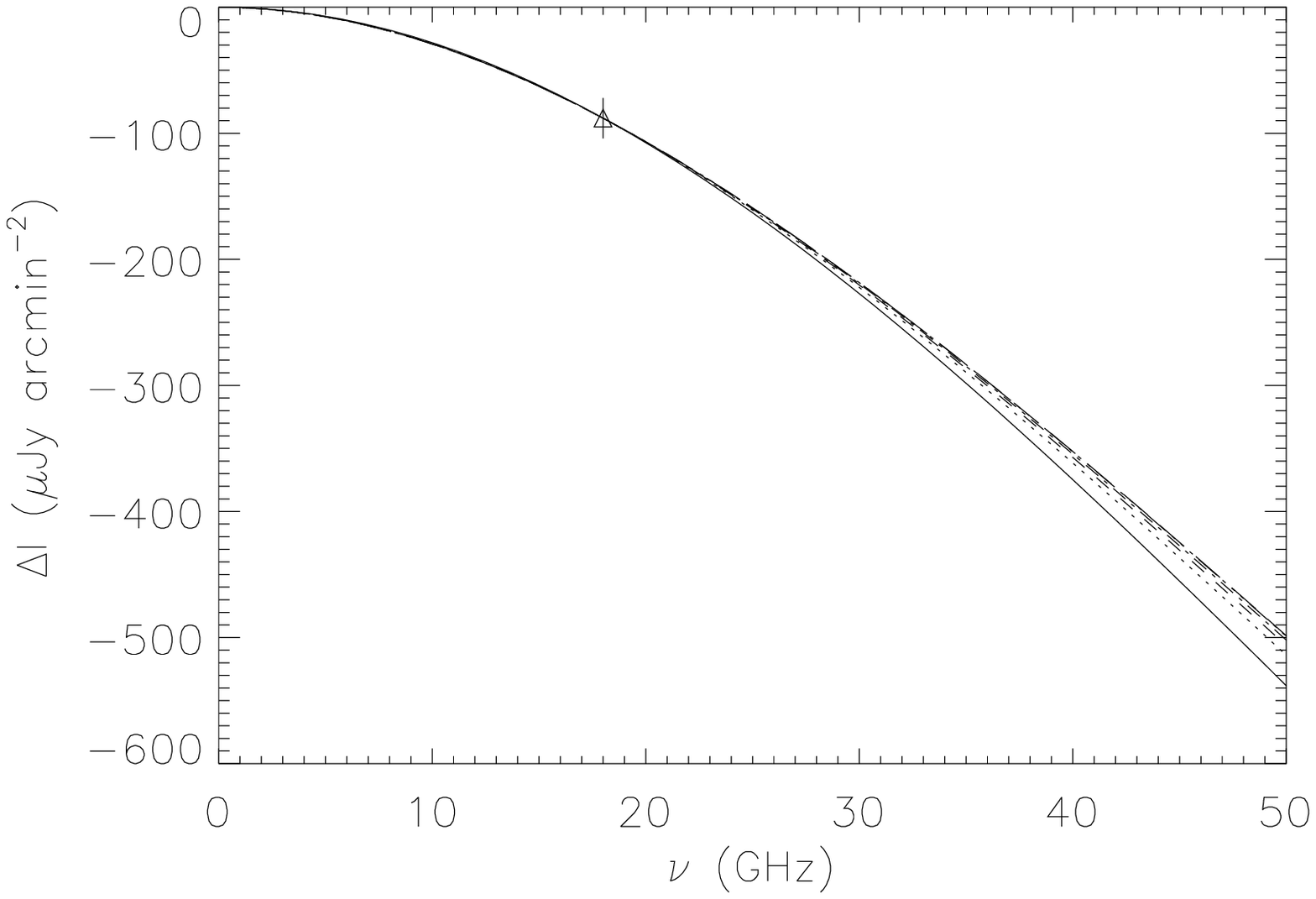}
	\includegraphics[width=\columnwidth,angle=0]{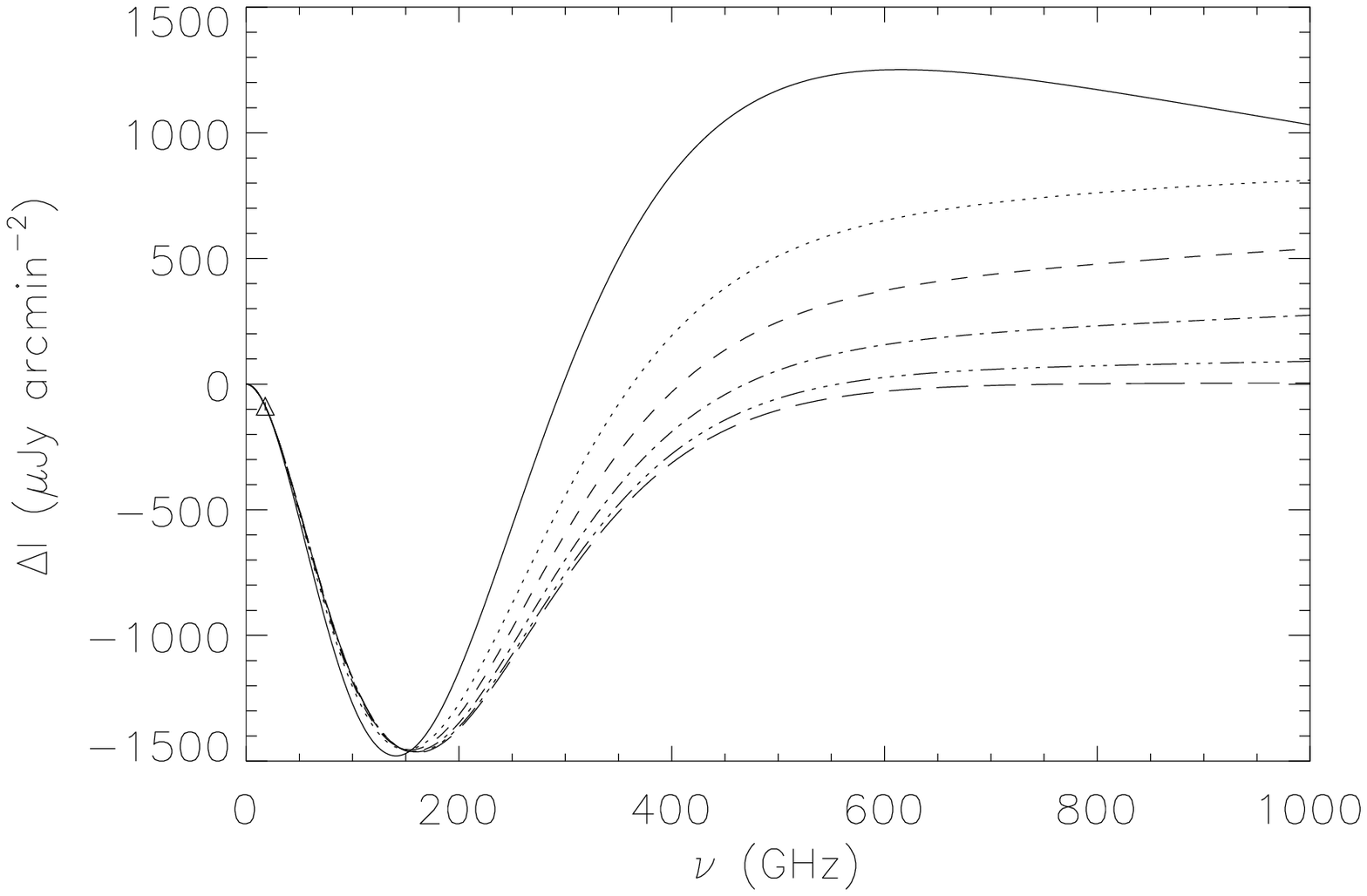}
	\includegraphics[width=\columnwidth,angle=0]{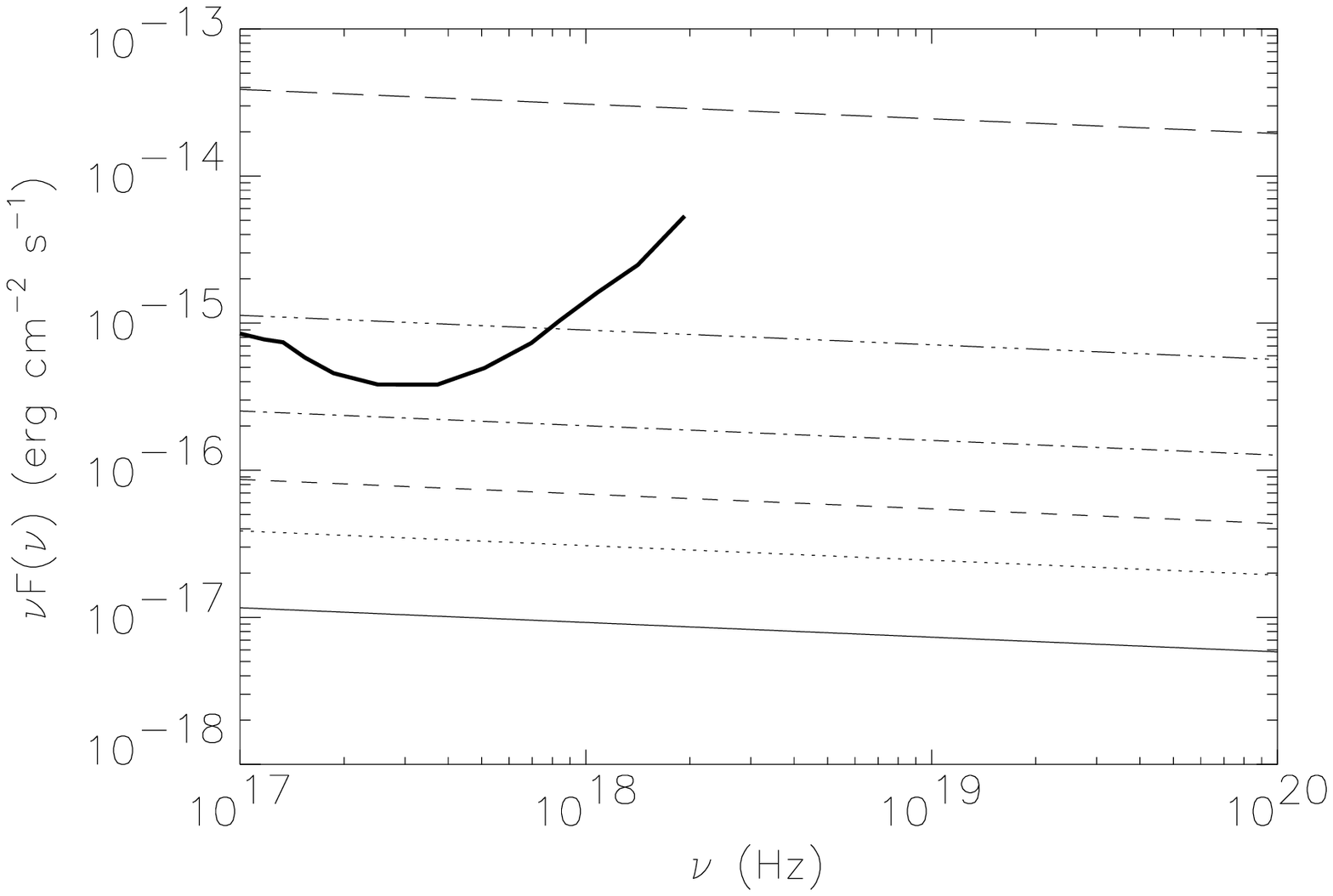}
    \caption{Spectrum of the \textbf{non-thermal} SZE (in surface brightness units) at low
      (upper panel) and high frequencies (middle panel) and of the ICS
      emission in the X-rays integrated inside a square with side
      length of 4 arcsec (lower panel) for an electrons non-thermal
      population with $s_e=3.2$. 
The following models are plotted: 
solid line: non-thermal SZE with minimum momentum of electrons $p_1=1$ and optical depth $\tau=7.6\times10^{-5}$; 
dotted line: $p_1=2$ and $\tau=5.5\times10^{-5}$;
dashed line: $p_1=3$ and $\tau=5.1\times10^{-5}$;
dot-dashed line: $p_1=5$ and $\tau=4.8\times10^{-5}$;
three dots-dashed line: $p_1=10$ and $\tau=4.7\times10^{-5}$;
long-dashed line: $p_1=50$ and $\tau=4.6\times10^{-5}$. 
The sensitivity of Chandra is plotted for an integration time of 100 ks \citep{2012SPIE.8443E..1ZT}.}
    \label{szspectrum_3p2}
%  \end{center}
\end{figure}

%{\large \bf Modeling the non--thermal SZ Effect}\\
We model the non-thermal SZE signal we detected in the RG01 lobe using a full relativistic formalism \citep{2003A&A...397...27C}, where the SZE spectrum is given by the expression
%\begin{equation}
$\Delta I(x)=I(x) - I_0(x)$.
%\label{eq.Deltax}
%\end{equation}
The incoming radiation spectrum is the CMB spectrum
%\begin{equation} \label{spettro_inc}
$I_{0}(x)=2\frac{(k T_0)^3}{(hc)^2} \frac{x^3}{e^x-1}$ ,
%\end{equation}
with $x = h \nu / k T_{0}$ and $T_0$ is the CMB temperature today. The resulting SZE spectrum is calculated according to the equation
%\begin{equation}
$I(x)=\int_{-\infty}^{+\infty} I_{0}(xe^{-s}) P(s) ds \;$,
%\label{spettro_risultante}
%\end{equation}
where $P(s)$ is the photon redistribution function (yielding the
probability of a logarithmic shift  $s=\ln (\nu'/\nu)$ in the photon
frequency due to the inverse Compton scattering process) that depends
on the electron spectrum producing the CMB Comptonization, and
includes all the relativistic corrections. It is calculated by the sum
of the probability functions to have $n$ scatterings, $P_n(s)$,
weighted by the corresponding Poissonian probability: 
%\begin{equation}
$P(s)=\sum_{n=0}^{+\infty} \frac{e^{-\tau} \tau^n}{n!} P_n(s) \;$,
%\end{equation}
where the optical depth $\tau$ is given by the integral along the line of sight $\ell$ of the electron density
%\begin {equation}
$\tau=\sigma_T \int n_e d\ell \;$ ,
%\end{equation}
where $n_e$ is the electron plasma density.
Each function $P_n(s)$ is given by the convolution product of $n$ single scattering probability functions 
$P_1(s)$:
\begin{equation}
P_n(s)=\underbrace{P_1(s) \otimes \ldots \otimes P_1(s)}_{\mbox{n times}} \mathrm{where~} P_1(s)=\int_0^\infty f_e(p) P_s(s,p) dp \; ,\nonumber
 \label{eq.pns}
\end{equation}
%where
%\begin{equation}
%P_1(s)=\int_0^\infty f_e(p) P_s(s,p) dp \;,
%\label{eq.p1s}
%\end{equation}
and where $f_e(p)$ is the electron momentum distribution function
(normalized as to have $\int_0^\infty f_e(p) dp=1$), and $P_s(s,p)$ is
the function that gives the probability to have a frequency shift $s$
by an electron with a-dimensional momentum $p=\beta \gamma$, and is
given by the physics of the inverse Compton scattering process (see,
e.g., \citet{2003A&A...397...27C,2000A&A...360..417E}).\\
%
%In the case of the thermal SZE produced by a thermal electron distribution with temperature $kT$, the function $f_e(p)$ has the following form:
%\begin{equation} \label{termica}
%f_e(p)=\frac{\eta}{K_2(\eta)}p^2 \exp\left(-\eta \sqrt{1+p^2}\right) \;,
%\end{equation}
%where $\eta=(m_e c^2)/(k T)$ and $K_2 (\eta)$ is the modified Bessel function of second kind \citep{1965hmfw.book.....A}.
For non-thermal electrons we use a single power-law electrons momentum distribution with a minimum momentum $p_1$ \\
%\begin{equation}
$f_e(p)\propto p^{-s_e} \; ; \;\;\;\; p_1\leq p\leq p_2 $, \\
%\end{equation}
and we assume a high value of the maximum momentum ($p_2=10^8$).\\
Our 18 GHz observation for the RG01 SZE signal can be fitted, in
principle, with both a thermal or a non-thermal electron population
with different values of the spectral index and the minimum momentum
of electrons (for the non-thermal SZE) and of the temperature (for the
thermal SZE), leaving the optical depth as a free parameter: this is,
in fact, the result of a degeneracy in the SZE parameters at
low-frequencies. However, the absence of any detectable X--ray
emission in this region, its large distance from the centre of
the cluster merger ($\approx$ 800 kpc), and the absence of any diffuse
emission in a region roughly 400 kpc region E to W from the
westernmost edge of X--ray emission, %(see Fig. \ref{4freqs} panel
                                %{\bf (d)}), 
imply that emission observed in this region does not have a thermal origin.\\
In order to break this parameter degeneracy, we obtained information
about the spectral index of the electrons in the RG01 lobe  from the
observed radio spectrum at frequencies in the range 2.1-9.0 GHz: the
average radio spectral index measured between 2.1 and 9 GHz is
$\alpha_r=1.1\pm0.15$. This corresponds to a range of electrons
spectral index values  in the range $2.9 \le s_e \le 3.5$, where
$s_e=2 \alpha_r +1$. The shape of the radio spectrum  indicates that
we are in the presence of a quite typical non-thermal electron
distribution in the RG01 lobe; this important fact allows us to
constrain the range of possible SZE models that can fit the observed
SZE signal at 18 GHz, thus restricting our analysis to non-thermal
models of the SZE. \\
Fig. \ref{szspectrum_3p2} reports the non-thermal SZE in the RG01 lobe
calculated with the value of the average radio spectral index
measured between 2.1 and 9 GHz  $\alpha_r=1.1$ corresponding to
$s_e=3.2$ and is shown at low frequencies ($\nu < 50$ GHz), at high
frequencies ($\nu < 1000$ GHz), as well as the ICS X-ray emission
expected for several values of the minimum momentum $p_1$ of the
non-thermal electron distribution. All the non-thermal SZE models which are consistent with the observed
radio spectrum of the RG01 lobe can fit the ATCA SZE signal at 18 GHz
confirming that the SZE signal we detected is of non-thermal origin
and it is produced by a non-thermal electron population whose energy
spectrum is consistent with the observed radio synchrotron spectrum.\\
This is the first detection of a non-thermal SZE effect in the lobe of a radio galaxy.
%
%The expectations of the high-frequency non-thermal SZE from the RG01
%lobe and the ICS X-ray emission from the same electron population can
%also be observed with mm and sub-mm experiments with appropriate
%sensitivity and angular resolution (like ALMA and Millimetron), as
%well as with the Chandra X-ray observatory, depending on the value of
%$p_1$.\\ 
%
The upper limit on X-ray emission from the RG01 lobes provided by
Chandra indicates an upper limit on the value of $p_1$ between 5 and 10
that correspond to minimum electron energies $E_{\rm min} < 2.5-5$ MeV. 
%
%The reason why these values are derived as upper limits is due to the
%fact that the SZE has been measured at low frequencies where its
%negative amplitude depends mainly on the electrons optical depth
%regardless of their energy (note, however, that there is a small
%dependence on the electrons energy that is appreciable from the
%slightly different values of the optical depth required for different
%value of $p_1$, as shown in Fig.\ref{szspectrum_3p2}). Therefore, if
%the value of $p_1$ is higher, it is necessary to assume that the
%electrons spectrum has a higher normalization to fit the SZE, and this
%in turn implies a higher number of high-energy electrons and a
%stronger ICS X-ray emission. 

%\vspace{-1em}
The central panel of Fig.3 shows that a more precise determination of
the value of $p_1$ can be obtained by measuring the spectrum of the
non-thermal SZE at high frequencies, i.e. at the crossover frequency,
that can vary between $\sim 250$ and $\sim 500$ GHz depending on the
value of $p_1$, or in the high energy part of the spectrum, where the
SZE is positive. This is in agreement with previous studies of the
properties of non-thermal electrons in radio galaxy lobes from SZE
measurements (see \citet{2011A&A...535A.108C}). We note that
a measurement of $p_1$ can lead to an estimation of
the optical depth $\tau$, and this value in turn leads to crucial information about
the physics of radio galaxy lobes; namely, density of non-thermal electrons and the intensity of the
magnetic field in the lobe (see discussion in \citet{2011A&A...535A.108C}).

\section{Conclusions}
%{\large \bf Conclusions}\\
The first detection of the non-thermal SZE presented here in a radio galaxy jet, after its theoretical
prediction (see \citet{2008MNRAS.385.2041C}), is a significant step towards the characterization of the low--energy cut--off in a non--thermal plasma, and
its usefulness and importance therefore cut across several fields in
astrophysics. Extension of the non-thermal SZE observed in the lobe of RG01 at higher frequencies can also be observed with mm and sub-mm
experiments with appropriate sensitivity and angular resolution (like
ALMA and Millimetron).\\ 
Future detection of X-ray emission from the same physical process, i.e. the up-scattering of CMB photons by non--thermal populations of electrons in the lobe of RG01, combined with radio and SZ Effect, will provide a value of the overall energy extension and spectral shape of the energy spectrum of these non-thermal electrons
residing in lobes of radio galaxies. 

%{\large \bf Acknowledgements}\\
\section*{Acknowledgements}
The Australia Telescope Compact Array is part of the Australia
Telescope which is funded by the Commonwealth of Australia for
operation as a National Facility managed by CSIRO. X-ray data was obtained from the Chandra Data Archive, observations made by the Chandra X-ray Observatory and published previously in cited articles. The radio / X--ray overlay image was made using the KARMA package \citep{1996ASPC..101...80G}. Observations and
analysis were made possible by a generous grant for Astronomy by IIT
Indore, and travel funding for S.M. by RRI, Bangalore and IIT Indore. S.C. acknowledges support by the South African Research Chairs
Initiative of the Department of Science and Technology and National
Research Foundation (Grant No 77948) and by the Square Kilometre Array
(SKA). P.M. acknowledges support from the DST/NRF SKA post-graduate
bursary initiative. 

%%%%%%%%%%%%%%%%%%%%%%%%%%%%%%%%%%%%%%%%%%%%%%%%%%

%%%%%%%%%%%%%%%%%%%% REFERENCES %%%%%%%%%%%%%%%%%%

% The best way to enter references is to use BibTeX:
\vspace{-2em}
\bibliographystyle{apj}
\bibliography{bullet7} 

%%%%%%%%%%%%%%%%%%%%%%%%%%%%%%%%%%%%%%%%%%%%%%%%%%

% Don't change these lines
%\bsp	% typesetting comment
\label{lastpage}
\end{document}